# Measurement of High-temperature Thermophysical Properties of Bulk and Coatings Using Modulated Photothermal Radiometry

Jian Zeng[1,*], Ka Man Chung[2,*], Qingyang Wang[1], Xiaoxin Wang[3], Yu Pei[1], Peiwen Li[3], Renkun Chen[1,2,#]

[1]Department of Mechanical and Aerospace Engineering, University of California San Diego, La Jolla, California 92093, United States

[2]Material Science and Engineering Program, University of California San Diego, La Jolla, California 92093, United States

[3]Department of Aerospace and Mechanical Engineering, The University of Arizona, Tucson, Arizona 85721, United States

[*]These authors contributed equally. [#]Corresponding author: rkchen@ucsd.edu

**Abstract**

This paper presents the development of instrumentation for the measurement of high-temperature thermal conductivity of bulk and coatings using a modulated photothermal radiometry (MPR) method, where a sample is heated by an intensity-modulated laser to probe into different layers of the sample. While MPR has been previously established, most of the previous studies only focus on the measurement at room temperature. The MPR has not been well studied for measurements of bulk and coating materials at high temperatures, which are increasingly important for a multitude of applications, such as materials used in the concentrating solar power (CSP) plants and the nuclear reactors. MPR is a non-contact technique that utilizes the intrinsic thermal emission from the specimens for thermometry, which is favorable for measurements at high temperatures in harsh environment. The authors designed and utilized a sample holder suitable for high temperature measurement up to 973 K with good temperature uniformity within the sample. The high-temperature MPR setup was validated by measuring bulk materials with known thermal conductivity. The setup and technique were then extended to the measurement of black solar-absorbing coatings of 10 to 50 μm thick on various substrates by modulating the frequency of the laser heating beam and the thermal penetration depth. The studies showed that thermal conductivities of typical solar-absorbing coatings are 0.4 ~ 0.8 W m$^{-1}$ K$^{-1}$, indicating a possibly large temperature drop within the coating under high solar irradiation flux, such as over 1000-sun for central solar towers in CSP plants.

## 1. Introduction

Thermal transport at high temperatures is a significant process for many applications [1], such as concentrated solar power (CSP) plants [2, 3], thermal energy storage [4], and thermal barrier coatings for gas-turbine engines [5]. There are several established techniques for high temperature thermal conductivity measurement, such as the laser flash analysis (LFA) [6, 7], transient hot-wire (THW) [8], hot-disk transient plane-source (TPS) [9, 10], 3ω method [11-13], as well as the pump-probe time or frequency domain thermoreflectance (TDTR/FDTR) techniques [14-17], etc. However, there are still limitations of these techniques under certain circumstances, for example, on samples with rough surfaces and at high temperature. In a typical LFA measurement, the thermal diffusivity could be obtained from the time delay of temperature rise at the backside of the sample upon pulsed heating. Although LFA is capable of measuring bulk materials with high throughput, it is insensitive to thin films [18, 19] and the typical commercial LFA instrument can only measure samples with a thickness of ~ 100 μm or above [20, 21]. The 3ω method [11] is another popular technique that is capable of measuring bulk materials and coatings with the thickness from 10 μm [22] down to 10-100 nm [23]. However, the 3ω technique needs a thin metallic strip serving both as a heater and a thermometer, and thus it requires time-consuming microfabrication process that is sometimes not feasible for rough surfaces [24]. Besides, it is challenging to preserve the functionalities of the 3ω heater at high temperature in air. TDTR/FDTR are noncontact optical heating and sensing techniques to measure the thermal properties of both bulk and thin films. Thin films can be accurately measured with a short thermal penetration depth via either short time delay or high modulation frequency [15, 16, 25]. However, the techniques also need a thin and optically smooth transducer layer (e.g., Au or Al) [26] for pump laser absorption and probe laser reflectance, which poses challenges at high temperature and for rough surfaces. The THW and TPS methods work well for rough surfaces, e.g., particles [27] and liquids [8], however, they rely on the electric resistance of heater/sensor wires (e.g., Pt wire for THW) in intimate contact with the sample, thus possessing a similar challenge in thermal stability as other contact methods.

Photothermal radiometry (PTR) using either modulated (continuous wave) and pulsed laser is a well-known non-destructive and non-contact thermal characterization tool for both bulk materials [28, 29] and thin film specimens [30-32]. In the modulated photothermal radiometry (MPR), the sample is heated by an intensity-modulated laser to probe into different depths of the sample [29, 33]. The MPR uses the thermal emission from the specimen surface as the thermometry probe. While this thermometry approach does not have the high spatial resolution offered by the thermoreflectance method, it is not affected by the stability issue of thermal transducers and surface roughness as in the 3ω method and the thermoreflectance techniques. Therefore, MPR can serve as an attractive and convenient alternative for measuring thermal conductivity of rough or porous samples at high temperature. While MPR has been previously established, most of the previous studies only focus on the measurement at room temperature [28, 29, 34] despite a few reports on the pulsed photothermal radiometry at the elevated temperature [35]. Due to the intrinsically non-linear relationship between the infrared (IR) emissive power ($E_a$) and temperature, i.e., $E_a \sim T_o^3$ for surface temperature oscillating around the baseline sample temperature of $T_o$, the emission signal increases substantially at higher temperatures for MPR, leading to a greater measurement sensitivity at higher temperature, which is advantageous over many other techniques. In the meantime, there are also perceived challenges associated with high-temperature MPR measurement, including a suitable sample holder with good temperature uniformity within the measurement spatial window and a suitable black laser absorption and IR emission coating suitable at high temperature in air.

In this study, we demonstrate the MPR measurement of bulk materials and coatings up to 973 K. The MPR setup is carefully designed for high temperature measurement with consideration of heat loss and temperature uniformity. We first calibrate the MPR system by measuring common bulk solids, including borosilicate glass [11, 36, 37], Pyroceram 9606 [38, 39], 8 mol% $Y_2O_3$ stabilized $ZrO_2$ (8YSZ) [40] and stainless steel 304 (SS304) [41], to confirm the acceptable measurement error from MPR (< 10%).

Subsequently, we measure several solar-absorbing coatings, including Pyromark 1200, Pyromark 2500, and black spinel oxide ($Cu_{0.5}Cr_{1.1}Mn_{1.4}O_4$) coatings synthesized before by the authors [42], with variation in the substrate materials and coating thickness. We find that the thermal conductivities for these coatings are in the range of 0.4 - 0.8 W m$^{-1}$ K$^{-1}$ from room temperature to 700 °C, depending on the compositions and temperature, indicating a possibly considerable temperature drop within the coatings for high-temperature CSPs. The MPR technique reported here can provide a facile thermal conductivity diagnostic tool for high temperature materials with broad applications, such as CSP, nuclear reactors, and thermal barrier coating for turbines, etc.

## 2. Experimental Section

### 2.1. Principle of MPR Measurement

**Figure 1** shows the principle of the MPR measurement on a two-layer sample. The sample is heated up by the intensity-modulated laser $q_s$, leading to the oscillation of the surface temperature rise $\theta_s$ of the sample at the same frequency ($f$) of the heating laser. Thermal emission from the sample surface is collected by and IR-detector which will be used to calculate the $\theta_s$. For a two-layer sample with thicknesses of $L_1$ and $L_2$ for the top layer and the substrate respectively, the laser flux is assumed to be absorbed at the surface of the first layer. The light absorbing coatings (10-50 μm) on the sample surfaces are based on Pyromark 1200, Pyromark 2500, and a black spinel oxide ($Cu_{0.5}Cr_{1.1}Mn_{1.4}O_4$) [42]. The coating has small skin depth (< 5 μm [43, 44] ) at the laser wavelength. Therefore, the assumption of the surface heat flux is valid. To simplify the problem to one-dimensional (1D) heat transfer, one has to make sure that the thermal penetration depth, $L_p = \sqrt{2\alpha/\omega}$, where $\alpha$ is the thermal diffusivity and $\omega$ is the angular frequency, is much smaller than the laser spot size, which is around 6 mm in this experiment. The thermal penetration depth can be adjusted by changing the modulation frequency of the incident laser. For example, at low frequency where $l_1 \ll L_p < l_1 + l_2$, the thermal response is mainly determined by the property of the second layer (i.e., substrate). We also control $L_p$ to be smaller than the total thickness of sample to avoid the backside effect (i.e., utilizing the adiabatic boundary condition on the backside). At high frequency where $L_p < l_1$, the thermal response is only sensitive to the first layer (i.e., coating). By modulating the thermal penetration depth, we can probe into different layers of the sample.

The 1D heat transfer model for a multi-layer sample with the adiabatic boundary condition on the back side and heating on the front side is [21, 45]:

$$\theta_s = -\frac{d}{c} q_s \tag{1}$$

where $\theta_s$ is the surface temperature oscillation defined as $\theta_s = T_s - T_o$; $T_s$ is the transient surface temperature and $T_o$ is the baseline surface temperature by the cartridge heaters and DC component of the laser heating; $q_s$ is the AC component of the laser heat flux; $d$ and $c$ are elements in the following transfer matrix:

$$M = \begin{pmatrix} a & b \\ c & d \end{pmatrix} = M_n M_{n-1} \cdots M_i \cdots M_1 \tag{2}$$

where $M_i$ is the transfer matrix for the $i^{th}$ layer with the subscript 1 for the top layer and $n$ for the bottom layer. For the two-layer sample shown in **Figure 1b,** the first layer is the coating and the second layer is the substrate. The transfer matrix $M_i$ is expressed as follows:

$$M_i = \begin{pmatrix} \cosh(D_i\sqrt{j\omega}) & -\frac{\sinh(D_i\sqrt{j\omega})}{e_i\sqrt{j\omega}} \\ -e_i\sqrt{j\omega}\sinh(D_i\sqrt{j\omega}) & \cosh(D_i\sqrt{j\omega}) \end{pmatrix} \tag{3}$$

where $D_i = \frac{l_i}{\sqrt{\alpha_i}}$, $l_i$ and $\alpha_i$ are the thickness and thermal diffusivity of the $i^{th}$ layer. $j$ is the imaginary. $e_i$ is the thermal effusivity of the $i^{th}$ layer.

In the MPR measurement of solar-absorbing coatings on a substrate, the coating absorbs the laser flux and emits thermal radiation. Therefore, the multi-layer model is reduced to the two-layer model assuming that the substrate is a semi-infinite medium (adiabatic boundary condition on the back side):

$$\theta_s = \frac{q_s}{e_f\sqrt{j\omega}} \left[ \frac{e_f \cosh(D_f\sqrt{j\omega}) + e_s \sinh(D_f\sqrt{j\omega})}{e_f \sinh(D_f\sqrt{j\omega}) + e_s \cosh(D_f\sqrt{j\omega})} \right] \tag{4}$$

where $e_f$ and $e_s$ are thermal effusivities of the coating and the substrate, respectively; $q_s$ is laser heat flux;

$D_f = \frac{l_f}{\sqrt{\alpha_f}}$, where $\alpha_f$ is the thermal diffusivity of the coating; $a_f = k_f/(\rho_f c_f)$, where $\rho_f$, $c_f$ and $k_f$ are the density, specific heat, and thermal conductivity of the coating, respectively. Notably, The measurement for bulk material is conducted at the low frequency range (1 Hz $< f <$ 20 Hz) with $L_p \gg l_f$, where $l_f$ is the thickness of the coating. Under this condition, **Equation 4** is reduced to **Equation 5** at the low frequency limit with $L_p \gg l_f$.

$$\theta_s = \frac{q_s \exp\left(-\frac{\pi}{4}i\right)}{e_s \sqrt{\omega}} + \frac{q_s l_f}{k_f} \tag{5}$$

where the first term on the right-hand side depends on the thermal effusivity of the substrate ($e_s = \sqrt{\rho_s c_s k_s}$) and is proportional to $\omega^{-1/2}$ while the second term depends on the properties of the coating and is frequency independent. To obtain the thermal properties of the coating ($e_f$ and $a_f$), the thickness of the coating and the laser flux were measured separately, which contributes to the measurement uncertainty, as we shall discuss in the **Section 3.1**. $k_f$ and $\rho_f c_f$ can be determined from the measured $e_f$ and $a_f$.

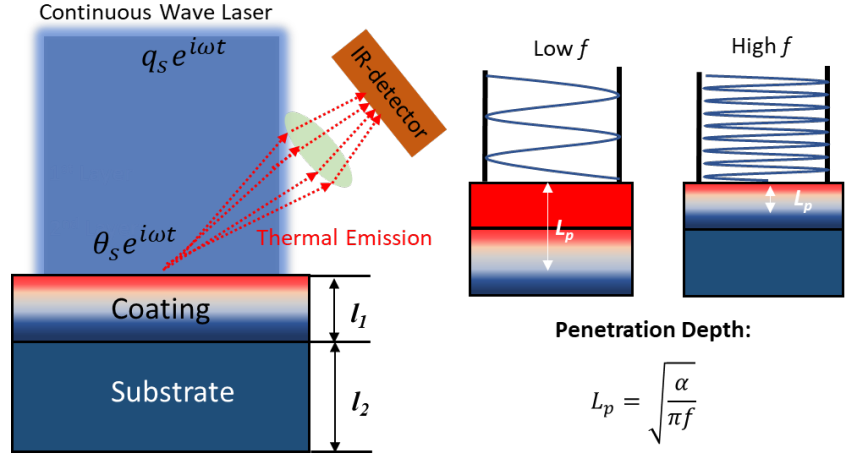

**Figure 1**. Principle of MPR measurement on a two-layer sample.

## 2.2. Details of MPR System

**Figure 2** shows the schematics of the MPR system and the principle of the measurement. As shown in **Figure 2a**, a continuous wave (CW) diode laser (Naku Technology Co., Ltd, CivilLaser, 445 nm, 6 W, frequency range: 0-30 kHz ,blue fiber-coupled laser) is driven by a waveform-generator (Agilent 33120A), providing a sinusoidal wave with intensity modulation at the frequency of $f$. Before heating the sample surface, the laser beam is homogenized to a top-hat profile using a homogenizer (EKSMA Optics, GTH-3.6-1.75FA). The laser spot size is controlled to be around 6 mm in diameter. The sample is heated up by the modulated laser flux, leading to the oscillation of the surface temperature rise $\theta_s$ of the sample at the same frequency ($f$) of the heating laser. Thermal emittance from the sample surface is collected with a pair of parabolic mirrors (Thorlab Inc., MPD169-G01, focal length of 3 inches) and focused to a high-speed HgCdTe (MCT) detector (Kolmar Technologies, Inc., KMPV11-0.25-J1/DC70). The IR detection spot size on the sample is estimated to be ~0.5 mm in diameter. The samples are coated with a thin black absorbing coating (10-50 μm) serving as a transducer layer with high visible absorptance at the laser wavelength (> 97%) and infrared thermal emittance (>90%) [42], as shown in **Figure 2a**. The coating is also stable up to 800 °C in air [42] for high-temperature operation. A germanium (Ge) filter (Thorlab Inc., WG91050-G) is placed in front of the MCT detector to filter short-wavelength light, such as the small reflection of the incident laser from the sample surface and the ambient light. The surface temperature rise $\theta_s$ is obtained from measuring the emitted power from the surface. The relationship between the measured voltage signal from the MCT detector and the surface temperature is calibrated using a pyrometer (Lumasense Technologies IGA 320/23-LO) that can also measure the temperature of the central area of 1 mm$^2$ within the laser spot. Thermal response signal (in voltage) from the MCT detector is recorded using a lock-in amplifier (Stanford Research System SR830) with the sinusoidal signal from the waveform-generator serving as the reference signal. The photograph of MPR system is shown in **Figure S1** in *Supplementary Information*.

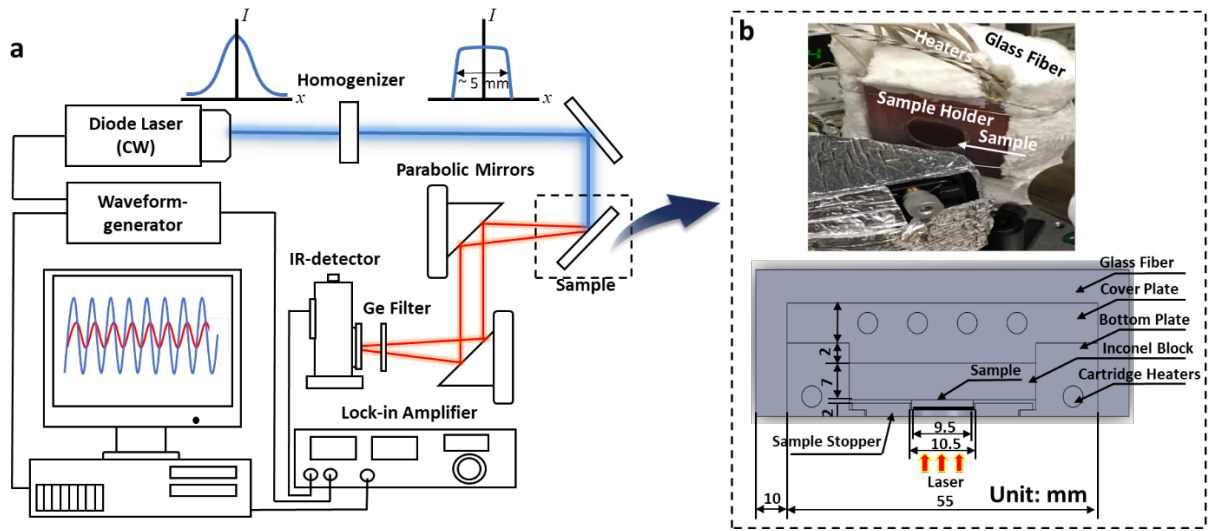

**Figure 2.** Details of MPR system. (a) Configuration of MPR system. (b) Photograph (top) and schematic (bottom) of the high-temperature sample holder for the MPR measurement. Samples are placed at the center of the holder made of Inconel 625 alloy.

A sample holder is designed for high temperature measurement up to 973 K in air. **Figure 2b** shows the photograph of the sample holder (top) and the cross-section view of the holder (bottom). The sample is made of a cover plate, a bottom plate, a block, and a sample stopper, all made of Inconel 625 alloy due to its high temperature stability in air. The cover plate and the bottom plate create a cavity to house the block and the disk-shaped sample of 10 mm in diameter. The sample is pressed by the sample stopper to make good contact with the underlying block. The sample stopper has a circular opening slightly smaller than the sample diameter for optical access. The entire setup is assembled using screws that tighten the cover plate and the bottom plate. The two plates are embedded with seven cartridge heaters of 6 mm in diameter providing a total power up to 420 W. The back and sides of the sample holder are wrapped by 10 mm thick glass fiber to reduce heat loss. The front surface of the sample is exposed for optical access.

The sample holder is designed to ensure the temperature uniformity in the sample. The temperature uniformity of the sample holder is verified by finite element modeling using COMSOL. **Figure 3a** shows the COMSOL simulation results for the temperature distribution inside the sample holder when the heater power is 170 W and the DC laser power is 2 W, which results in sample surface temperature $T_a$ of around 1050-1100 K. It is noted that the DC laser power, set at 2 W in this work, was inevitable to avoid the truncation of the AC component of laser power in our measurement. The laser spot size is set to be 6 mm in diameter. Both the convective and radiative heat losses from the surface are considered. The details of the COMSOL simulation are described in the ***Supplementary Information Section S4***. **Figure 3b and 3c** show the simulated temperature distribution within the sample ($SiO_2$) with diameter of 10 mm and thickness of 1.5 mm. **Figure 3b** shows the temperature distribution along the thickness direction at the center of the sample. Due to the heat loss from the sample surface and the fact that the heating power is applied from the sides and the back of the sample, there is a temperature gradient along the thickness direction. In our measurement, the sample surface was coated with a black light absorbing coatings for laser absorption and thermal emission [42], which led to a considerable radiative heat loss. Temperature gradient within the Inconel block is smaller than that within the sample due to the higher thermal conductivity of Inconel. Since some of the heat loss is compensated by the DC component of the laser heating (2 W), the maximum temperature difference within the sample is about 10 K. In actual measurements, we only consider the temperature uniformity within the thermal penetration depth, which is generally less than 0.5 mm at 1 Hz, the lowest frequency used in our measurements (e.g., for glass with thermal diffusivity of ~ 0.6 $mm^2\,s^{-1}$, the penetration depth is ~ 0.3 mm at 1 Hz). Within this penetration depth, the temperature difference is less than 5 K, indicating negligible temperature non-uniformity along the thickness direction. While the DC component of the laser power is set to be 2 W throughout our measurement, the temperature non-uniformity can be further reduced with higher DC laser power to compensate the heat loss as shown in the ***Supplementary Information Section S4***. **Figure 3c** shows the temperature distribution on the surface of the sample along the lateral direction. In general, the temperature decreases from the center to the edge due to the laser heating effect until it is 4 mm away from the center where the sample is mainly heated up by the cartridge heaters. The maximum temperature difference along the lateral direction is ~ 45 K. In the actual measurement, the MPR detector focuses on the center of the sample with the detection spot of about 0.5 mm in diameter. Within the IR detection spot, the temperature difference is less than 5 K, indicating negligible temperature difference along the lateral direction. Overall, the above analysis indicates that the temperature difference is less than 5 K, along both the lateral and normal directions, within the sample up to 973 K in the frequency range of interest (i.e., > 1 Hz).

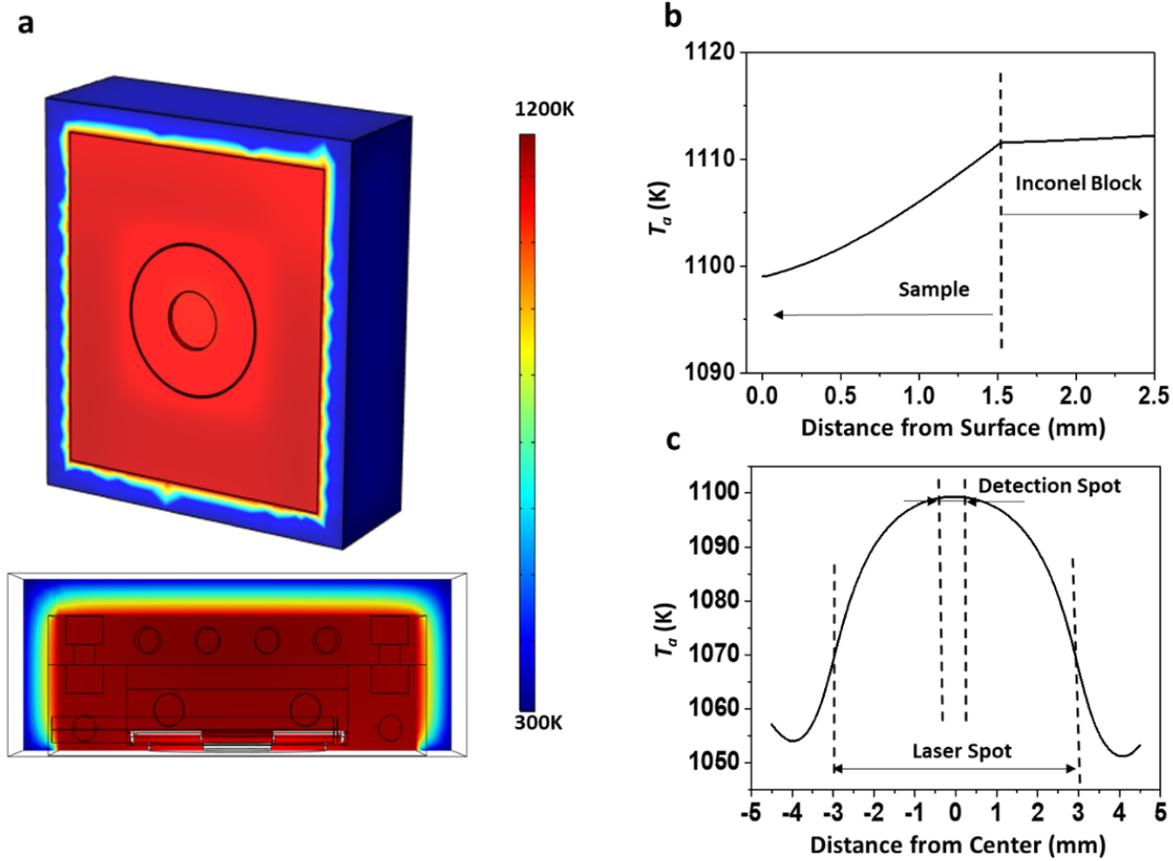

**Figure 3.** Finite-element simulation using COMSOL for the temperature distribution in the high-temperature sample holder. (a) Temperature profile of the sample holder with the heater power of 170 W and the DC component of laser power of 2 W. The upper row shows the external surface while the lower row shows the cross-section of the holder. (b) Temperature distribution within the sample along the thickness direction. (c) Temperature distribution along the radial direction on the sample surface.

### 2.3. Measurement of Bulk Materials

According to the **Equation 5**, we can obtain $e_s$ from the slope ($m_s$) of the $|\theta_s|$ vs $\omega^{-\frac{1}{2}}$ plot, i.e.:

$$m_s = \frac{d|\theta_s|}{d\omega^{-1/2}} \tag{6}$$

Since $m_s$ is only proportional to $e_s^{-1}$ for the same $q_s$, we can obtain the thermal effusivity of an unknown sample by comparing it to a known reference sample without measuring the heat flux and calibrating the surface temperature. This reduces the measurement uncertainty if the reference sample is well-known [15], because there is no need to measure the laser power and spot size separately. In this case, the thermal effusivity of the unknown sample could be obtained by:

$$e_s = \frac{m_{ref}}{m_s} e_{ref} \qquad (7)$$

where $e_s$ and $e_{ref}$ are denoted as the thermal effusivities of the unknown sample and the reference sample respectively; $m_s$ and $m_{ref}$ are the slope of $|\theta_s|$ vs $\omega^{-1/2}$ curves for the unknown sample and the reference sample, respectively. We also separately measured the thermal diffusivities of Pyroceram 9606 from 297K to 973K using a laser flash analyzer (NETSCH LFA 467), which showed very close results compared to Refs. [38, 39]. We used the specific heat data from Refs. [38, 39]. The obtained thermal effusivity for Pyroceram 9606 is shown in **Figure S7 in *Supplementary Information***.

## 2.4. Measurement of Solar-Absorbing Coatings on Substrates

We demonstrate the high temperature MPR measurement of coatings by measuring the solar-absorbing coatings, including Pyromark 1200, Pyromark 2500, and black spinel oxide ($Cu_{0.5}Cr_{1.1}Mn_{1.4}O_4$), all of which are suitable for high temperature CSP. Both the Pyromark 1200 and Pyromark 2500 (Tempil Co., USA) are state-of-the art solar-absorbing coatings for commercial CSP solar towers. They are silicone-based paints with high solar absorptance and good stability at high temperature in air. The $Cu_{0.5}Cr_{1.1}Mn_{1.4}O_4$ coating is made from black spinel oxide nanopowder, as reported by the authors previously [42]. We prepared Pyromark and black spinel oxide coatings following the same procedures reported in Refs.[42, 46]. Briefly, the Pyromark/black spinel oxide paints were dissolved in a mixture of organic solvents (xylene, toluene and isobutanol) and a resin (SILIKOPHEN), and were then spray-coated onto different substrates with controlled thickness. Subsequently, all the samples were annealed in a furnace up to 600 °C for curing. All the chemicals are procured from commercial providers, as reported in Ref. [42]. Different from the bulk material measurements, we sweep the modulation frequency from 1 to > 1000 Hz and fit the thermal response curve with **Equation 4** to directly obtain both $e_f$ and $D_f$. Since $D_f$ is related to the thermal diffusivity by $D_f = \frac{l_f}{\sqrt{\alpha_f}}$, $\alpha_f$ can be obtained with the knowledge of the coating thickness $l_f$. Subsequently, the $e_f$ and $\alpha_f$ will be used to calculate the thermal conductivity and volumetric heat capacity of the coatings. We also validate the fitting results for thermal conductivity using the intercept in the low frequency limit (**Equation 5**) and the phase signal as shown in the **Supplementary Information section S11**.

## 2.5. Sensitivity Analysis

**Figure 4** shows the sensitivity analysis, simulated for the MPR measurement on a 40 μm thick Pyromark 1200 coating on a SS304 substrate, with the following simulation parameters: $e_f = 900$ J s$^{-1/2}$ m$^{-2}$ K$^{-1}$, $a_f = 0.3$ mm$^2$ s$^{-1}$, and $e_s = 8703$ J s$^{-1/2}$ m$^{-2}$ K$^{-1}$. As shown in **Figure 4a**, the thermal response can be delineated into three regions, i.e., i) the substrate-dominated region in the low frequency range with $L_p > 100$ μm, where the thermal response is dominated by the properties of the substrate, indicated by the small slope; ii) the transitional region in the intermedium frequency range with 40 μm $< L_p < 100$ μm, where the thermal response is sensitive to both the substrate and the coating its slope changes rapidly; and iii) the coating-dominated region in the high frequency limit with $L_p < 40$ μm, where the thermal response is dominated by the properties of the coating, as indicated by the large slope. The phase of the thermal response as shown in **Figure 4b** deviates from 45° at the low frequency due to the existence of the coating but converges to 45° at high frequency, where the penetration depth is much smaller than the coating thickness and the heat transfer model is reduced to that in a homogeneous material (i.e., the coating). **Figure 4c** displays the sensitivity of the amplitude of thermal response to $e_f$, $\alpha_f$ and $e_s$, respectively. The sensitivity of $|\theta_s|$ is defined as:

$$S_\theta = \frac{\Delta|\theta_s|/|\theta_s|}{\Delta\gamma/\gamma} \tag{8}$$

where $|\theta_s|$ is the amplitude of surface temperature oscillation, $\gamma$ is the thermal property, which can be $e_f$,

$\alpha_f$ or $e_s$. For example, the sensitivity on $e_f$ of 1 means that 1% change in $e_f$ leads to 1% change in $|\theta_s|$.

**Figure 4d** shows the sensitivity of the slope $m$ of the $|\theta_s|$ vs $\omega^{-\frac{1}{2}}$ curve to $e_f$, $\alpha_f$ and $e_s$, respectively. The sensitivity of $m$ is defined as:

$$S_m = \frac{\Delta m/m}{\Delta \gamma/\gamma} \tag{9}$$

As expected, **Figures 4c and 4d** show that both the $|\theta_s|$ and $m$ are sensitive to $e_s$ in the low frequency limit (long thermal penetration depth) and to $e_f$ in the high frequency limit (short thermal penetration depth), while they are most sensitive to $\alpha_f$ within the transitional frequency range, because $\alpha_f$ directly dictates the thermal penetration depth.

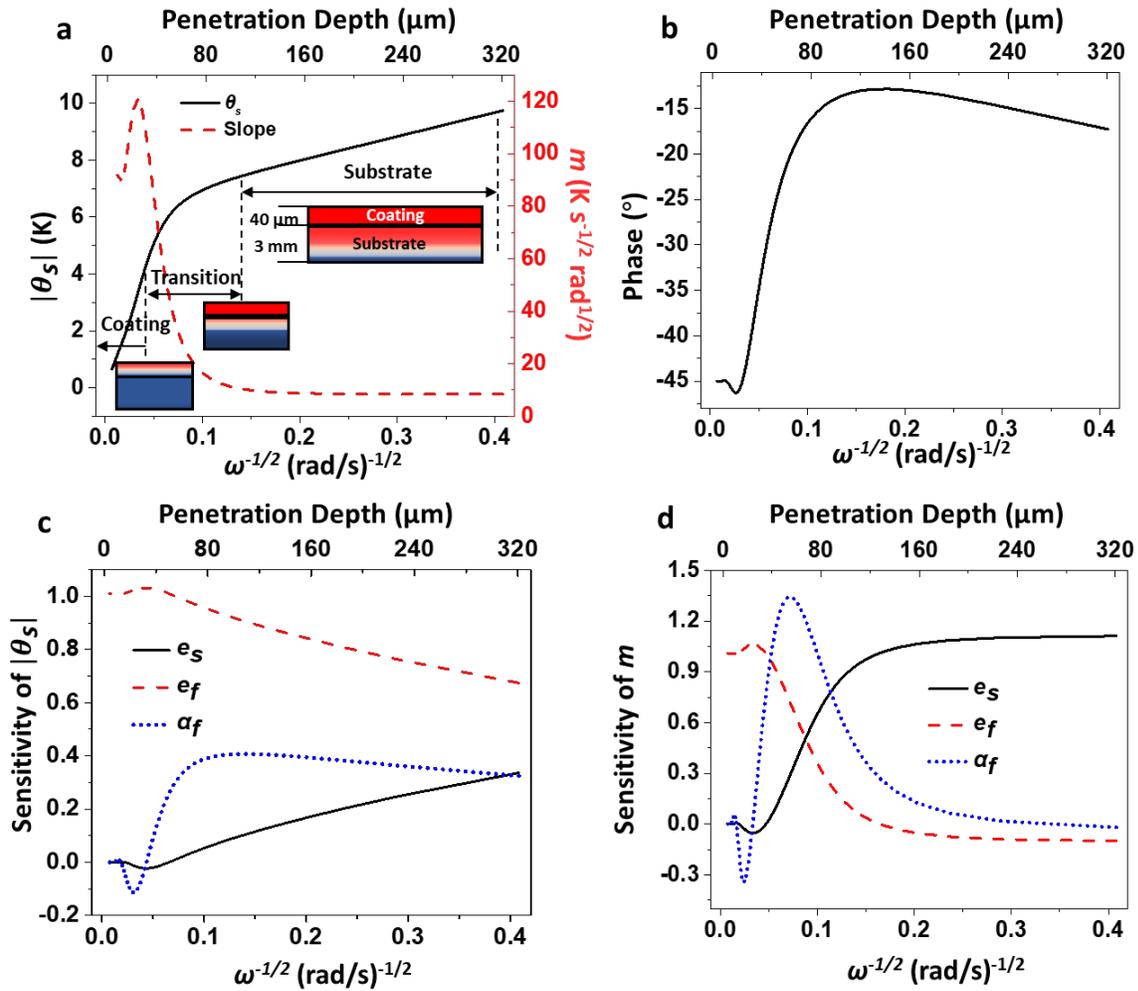

**Figure 4**. Simulation of the thermal response and the sensitivity analysis for a 40 μm Pyromark 1200

coating on a SS304 semi-infinite substrate. (a) $|\theta_s|$ (left y-axis) and slope $m$ of the $|\theta_s|$ vs $\omega^{-1/2}$ curve (right y-axis) as a function of $\omega^{-1/2}$ (bottom x-axis), and the corresponding thermal penetration depth in the coating ($L_p = \sqrt{2\alpha_f/\omega}$, top x-axis)). (b) Phase of thermal response. (c) Sensitivity of $|\theta_s|$ on $e_f$, $\alpha_f$ and $e_s$. (d) Sensitivity of $m$ on $e_f$, $\alpha_f$ and $e_s$.

## 3. Results and Discussion

### 3.1. Systematic Error of MPR Measurements

This section presents the analysis of systematic error of the MPR measurement. Typical measurement uncertainty of MPR is found to be ~ 10%, validating that the MPR technique is a feasible alternative for high temperature measurement and for samples with rough surface. The measurement uncertainty including both the systematic and random error which is the standard deviation of repeated tests. (**see section 3.2**).

**(a) Systematic Error of Measurement of Bulk Materials**

Thermal effusivity of bulk materials are measured by comparing the slope $m$ of thermal response curve $|\theta_s|$ vs $\omega^{-1/2}$ to a reference sample as shown in **Equation 6**. In this study, we used Pyroceram 9606 as the reference sample because its thermal properties were well-known with a small scattering in literature values of < 3% [38, 39]. Although the emissive power from the sample surface is intrinsically non-linear to $|\theta_s|$, it can be assumed linear within a narrow range of $|\theta_s|$, i.e., amplitude of the surface temperature rise due to the AC laser heating. Since the output voltage $V_{IR}$ of the IR detector is proportional to the received emissive power, it can be expressed as:

$$V_{IR} \propto 4\varepsilon\sigma T_o^3 |\theta_s| \qquad (10)$$

where $\varepsilon$ is the thermal emittance of sample surface, $\sigma$ is the Stefan-Boltzmann constant. Therefore, instead of comparing the slopes of the $|\theta_s|$ vs $\omega^{-1/2}$ curves, we can directly compare the slopes of the $V_{IR}$ vs $\omega^{-1/2}$ curves without doing the calibration for the IR detector, namely,

$$e_s = \frac{\tilde{m}_{ref}}{\tilde{m}_s} e_{ref} \tag{11}$$

where $\tilde{m}$ is the slope of $V_{IR}$ vs $\omega^{-\frac{1}{2}}$ curve (as opposed to $m$, the slope of $|\theta_s|$ vs $\omega^{-\frac{1}{2}}$ in **Equation 5**). By comparing the $\tilde{m}$ between the unknown sample and the reference sample, we can obtain $e_s$ without calibrating the IR detector or knowing $|\theta_s|$. The systematic measurement error therefore depends on the measurement accuracy of MCT (~ 3% due to the fluctuation of the ambient temperature and instrumentation noise) and the accuracy of the $e_{ref}$ for the reference Pyroceram 9606 sample (~ 3%) [38, 39]. Since both the $\tilde{m}_{ref}$ and $\tilde{m}_s$ are linear to the MCT signal, their uncertainties are also estimated to be ~ 3%. The systematic measurement error of bulk materials using MPR is calculated to be ~ 5.8% by the error propagation shown in **Equation 12**. To obtain the overall measurement uncertainty, the random error also needs to be considered, which will be discussed in **section 3.2**.

$$\frac{\Delta e_s}{e_s} = \sqrt{2\left(\frac{\Delta \tilde{m}}{\tilde{m}}\right)^2 + \left(\frac{\Delta e_{ref}}{e_{ref}}\right)^2} \tag{12}$$

**(b) Systematic Error of Measurement of Coatings**

The thermal effusivity and thermal diffusivity of coating are obtained by fitting the thermal response curve $|\theta_s|$ vs $\omega^{-\frac{1}{2}}$ from 1 Hz to 2 kHz using the least mean square method where the coefficient of determination, or $R^2$, was maximized. Different from the measurements of bulk materials, both the absolute value of the temperature and the laser heat flux are used for the data fitting. Therefore, the MCT detector was calibrated using a pyrometer in advance. The pyrometer could respond to the temperature oscillation below 20 Hz without waveform distortion; while at higher frequency (e.g., > 100 Hz), the signal would be distorted due to the long response time of the pyrometer. Therefore, $\theta_{s,exp}$ at high frequency was obtained by extrapolating the calibration curve obtained from 1 – 12 Hz. The calibration procedure is shown in detail in the ***Supplementary Information*** (**Section S7** and **Figure S8).**

At the low frequency limit, the thermal conductivity of the coating can also be obtained from **Equation 5** as proven in **supplementary information section S11,** where the measurement uncertainty is

given by:

$$\frac{\Delta k_f}{k_f} = \sqrt{\left(\frac{\Delta q_s}{q_s}\right)^2 + \left(\frac{\Delta l_f}{l_f}\right)^2 + \left(\frac{\Delta e_s}{e_s}\right)^2} \tag{13}$$

To obtain the thermal properties of the coating ($e_f$ and $\alpha_f$), the thickness of coating and the laser flux were measured separately, which contributed to the measurement uncertainty. Although the laser power could be accurately measured with an integrating sphere, it is much more difficult to measure the exact laser spot size accurately, leading to a large uncertainty in the laser heat flux [15]. Therefore, the laser heat flux was calibrated by the measurement of Pyroceram 9606 as a reference sample using a pyrometer as shown in **Figure 5a**. Since the laser flux is linear to the thermal effusivity of the reference sample, the uncertainty of laser heat flux is also 3%, the same as the measurement for bulk materials. The uncertainty of the coating thickness is 4.4% due to the porous and non-uniform nature of the coating introduced from the spray-coating process, and the uncertainty of the thermal effusivity of the substrate is 5.8% as shown in the **Section 3.1**

The systematic error of the thermal conductivity measurement for the coating is calculated to be 7.9% from **Equation 13**. At the high frequency limit where the penetration depth is much smaller than the coating thickness, the thermal response is the same as that for a bulk material with the coating properties, and the systematic error of thermal effusivity measurement of coating is 5.8%.

### 3.2. Validation of MPR on Bulk Materials

In this section, we validate the MPR measurement on bulk materials using different known materials with a wide range of thermal conductivities (e.g., borosilicate glass with $k$ of 1 ~ 2 W m$^{-1}$ K$^{-1}$ and SS304 with $k$ of 17 ~ 30 W m$^{-1}$ K$^{-1}$.) We used the Pyroceram 9606 as the reference sample to measure the thermal effusivities of other samples because of the consistent results reported in the literature [38, 39]. It is noted that all the bulk materials are coated with a black coating, however, only data at the low frequency (i.e., at long thermal penetration depth) are shown. **Figure 5a** shows the thermal response of Pyroceram 9606 at 473 K. According to **Equation 5**, the laser flux $q_S$ can be directly obtained from the slope of the curve with the knowledge of the thermal effusivity for Pyroceram 9606, $e_S$ = 2953.9 J s$^{-1/2}$ m$^{-2}$ K$^{-1}$ at 473 K, which is 82800 W m$^{-2}$. **Figure 5b** shows the repeatability of MPR measurement on Pyroceram 9606 for five tests. We spray-coated the Pyroceram 9606 with a 39 μm thick Pyromark 1200 for the first four measurements and an 18 μm thick black spinel oxide coating in the fifth measurement. Both coatings have similar absorptance at the laser wavelength and in the IR spectral range [42]. The same sample is unloaded and then loaded again between consecutive measurements, and thus the random error includes the possible minor changes in the position of the sample, position of the laser spot, and the view factor from the heated sample surface to the IR detector. For substrates with different coatings, the subsequent coating was made on the same substrate after removing the previous one. It demonstrates that the slopes for all the five measurements are consistent at the same temperature, indicating the same thermal diffusivity of the Pyroceram 9606 substrates (**Equation 5**). On the other hand, the intercepts at the *y* axis for the first four measurements are higher than that for the fifth measurement at the same temperature because of the larger coating thickness and the associated thermal resistance. **Table 1** shows that the random error of the MPR measurement for bulk materials, determined from the standard deviation of multiple measurements, is less than 3% at representative temperatures of 473 K, 673 K, and 973 K. The raw data for the repeated measurements at 973 K is shown **in Figure S11** in *Supplementary Information.* To estimate the measurement uncertainty, both the random (< 3%) and systematic errors (5.8% for thermal effusivity of bulk materials and 7.9% for thermal conductivity of coating) should be considered. The total measurement uncertainty of thermal effusivity of bulk materials is thus 6.5% and that of thermal conductivity of coatings is 8.5%. The error bars shown in the figures in following sections represent the measurement uncertainty including both the systematic and random errors.

**Table.1.** Random error of MPR measurement on Pyroceram 9606 with different coatings

| $T_O$ (K) | Slope $\widetilde{m}$ (mV rad$^{1/2}$ s$^{-1/2}$) | | | | | | Avg. | Std. | Random Error (%) |
|---|---|---|---|---|---|---|---|---|---|
| | 39 μm Pyromark 1200 | | | | 18 μm Black Spinel Oxide | | | | |
| | #1 | #2 | #3 | #4 | #5 | #6 | | | |
| 473 | 2.17 | 2.21 | 2.32 | 2.29 | 2.28 | - | 2.26 | 0.0622 | 2.76 |
| 673 | 6.41 | 6.52 | 6.69 | 6.55 | 6.54 | - | 6.54 | 0.0997 | 1.52 |
| 973 | 16.35 | - | - | - | 16.21 | 15.48 | 16.01 | 0.4645 | 2.90 |

Another fact worth noting in **Figure 5b** is that the slope at $T_o$ = 673 K is higher than that at 473 K because of the increasing emissive power at higher temperature. As shown in **Figure 5c**, the signal increases monotonically with increasing sample temperature from 473 K to 973 K. ***Supplementary Information Table S1*** summarizes the slope $\widetilde{m}$ of the $V_{IR}$ vs. $\omega^{-1/2}$ curves for Pyroceram 9606, which increases from 2.17 mV rad$^{1/2}$ s$^{-1/2}$ to 16.4 mV rad$^{1/2}$ s$^{-1/2}$ as $T_o$ increases from 473 K to 973 K. This is an intrinsic advantage of the photothermal radiometry where IR detection signal increases with $T_o^3$ (**Equation 10**), as shown in **Figure 5d**. Raw data for other bulk materials measured in this work are all included in ***Supplementary Information Section S2***, which also shows an increasing signal with increasing sample temperature.

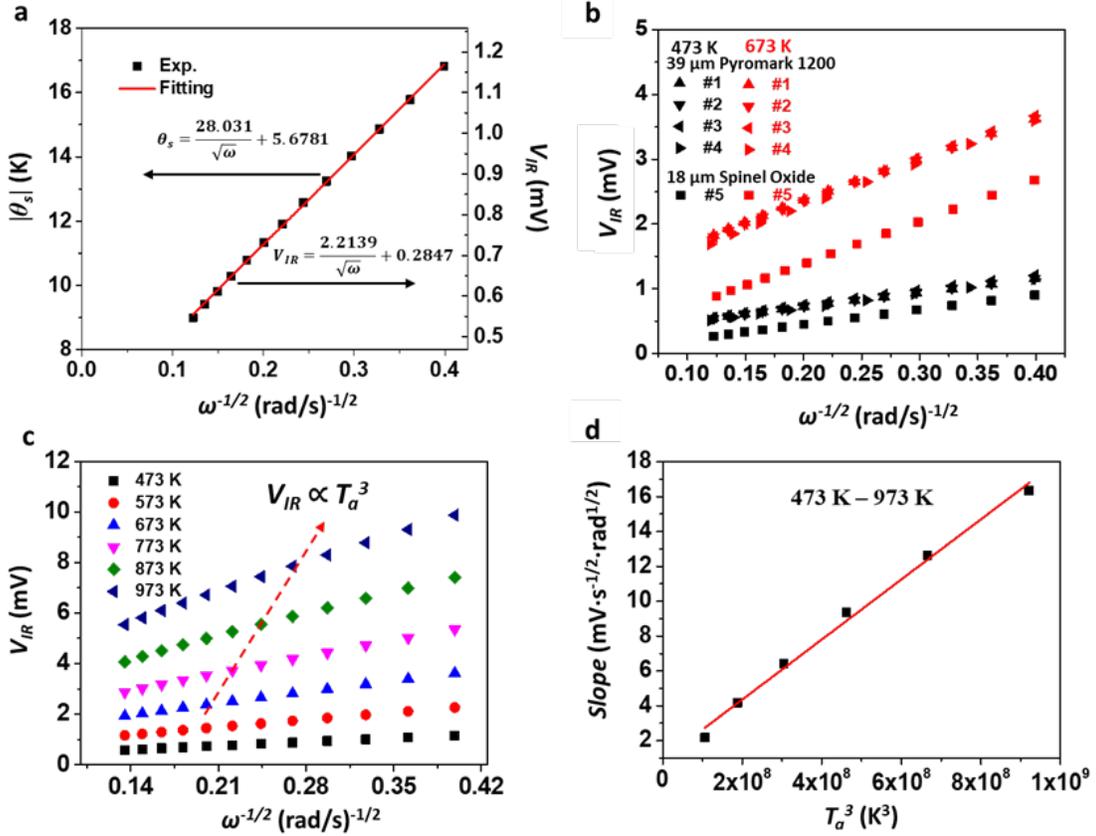

**Figure 5**. Validation of MPR measurement of bulk materials. (a) Measurement of laser heat flux using Pyroceram 9606 at 473 K as a reference sample. (b) Repeatability of MPR measurement and effect of coating materials. All the measurements were done on Pyroceram 9606 substrates. (c) IR detection signal of MPR measurement on a Pyroceram 9606 sample for sample surface baseline temperature ($T_o$) ranging from 473 K to 973 K. (d) Slope $\widetilde{m}$ of $V_{IR}$ vs $\omega^{-1/2}$ as a function of $T_o^3$, following **Equation 10**.

**Figure 6** compares the MPR measurement results of bulk materials to the reference values. We measured materials with well-established literature values: borosilicate glass [36, 37], Pyroceram 9606 [38, 39], 8YSZ [40], and SS304 [41]. The measured materials cover a wide range of thermal effusivity (1800 to 11000 J s$^{-1/2}$ m$^{-2}$ K$^{-1}$) and thermal conductivity (1.3 to 25 W m$^{-1}$ K$^{-1}$). We also measured a spinel oxide bulk pellet (Co$_{0.6}$Cr$_{0.6}$Fe$_{0.6}$Mn$_{0.6}$Ni$_{0.6}$)O$_4$. The spinel oxide powders were prepared by hydrothermal nanoparticle synthesis used by the authors previously [46] and then hot-pressed into a bulk pellet with over 95% relative density. **Figure 6a** compares the measured thermal effusivities to the literature or known values. The thermal effusivity is converted to thermal conductivity using the known specific heat and density values obtained from the literature [36-41] for glass, SS304, 8YSZ. For the spinel oxide, the reference thermal effusivity and thermal conductivity were obtained from thermal diffusivity measurement using the laser flash analyzer and the theoretically estimated specific heat (***Supplementary Information Figure S10***), as shown in **Figure 6b**. **Figure 6c** and **6d** show that the experimental results for both the thermal effusivity and thermal conductivity of the bulk materials are within 10% deviation from the reference values, which confirms the validity of the MPR technique for high temperature measurement.

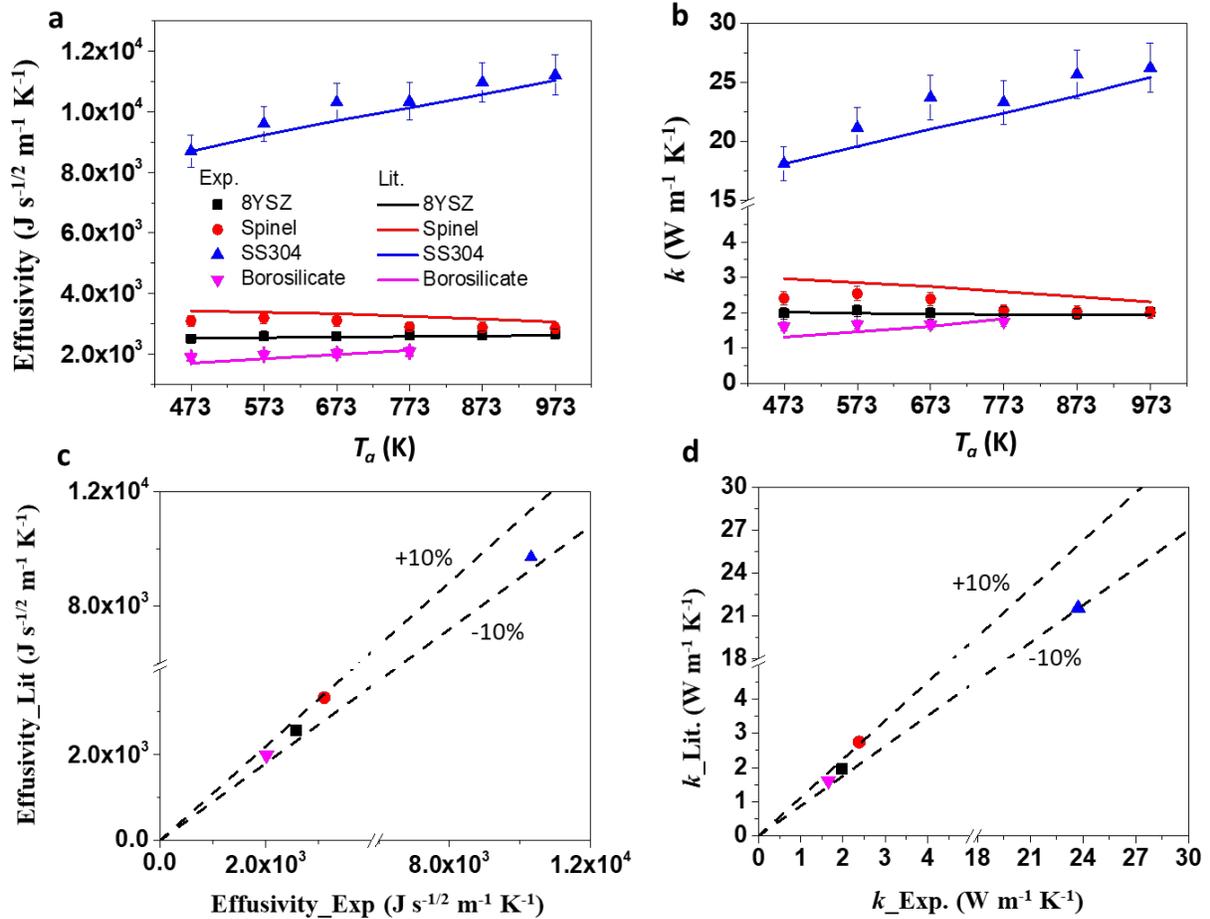

**Figure 6.** Validation of high temperature MPR measurement of bulk materials.(a) Measured thermal effusivity of different bulk materials up to 973 K. (b) Extracted thermal conductivity of different bulk materials up to 973 K by relating to the thermal effusivity using the $\rho c$ values. (c)

Comparison of the measured thermal effusivity to that from literatures or known values at $T_0 = 673\ K$. (d) Comparison of the measured thermal conductivity to that from literatures or known values at $T_0 = 673\ K$.

### 3.3. Measurement of Coating on Substrate

**Figure 7** shows the MPR measurement on a 53 μm thick Pyromark 1200 coating sprayed on an 8YSZ substrate. **Figure 7a** displays the surface profile of the Pyromark 1200 coating measured using a DektakXT stylus profilometer. The thickness of the coatings was also measured using the profilometer. The surface roughness of the coating is measured to be ~ 3 μm due to the porous nature and the build-up of the paint at the edge of coating during the spray-coating process. However, during the MPR measurement, the laser spot (~ 6 mm diameter) is focused at the center of the sample and thus the uncertainty of the coating thickness is less than 3 μm. **Figure 7b** depicts the thermal response of the coating on the 8YSZ substrate at different temperatures. Notably, the slope of the curve is smaller in the low frequency range where the signal is dominated by the thermal properties of the 8YSZ substrate. We can extract the thermal properties of the substrate using this range of frequency as shown in **Section 3.2**. In the high frequency range where the penetration depth is much smaller than the coating thickness, the thermal response is dominated by the properties of the Pyromark 1200 coating. The slope is larger in this range of frequency, indicating a lower thermal effusivity of Pyromark 1200 compared to that of 8YSZ. The transitional penetration depth, where the slope changes rapidly, is about ~ 60 μm, which is close to the thickness of the coating. The transitional frequency is therefore dependent on the thermal diffusivity of the coating. Therefore, we can also obtain the thermal effusivity and diffusivity of the coating by fitting the entire $|\theta_s|\ vs\ \omega^{-1/2}$ curves according to **Equation 4** using the least mean square algorithm, with the information of the surface heat flux ($q_s$ = 82800 W m$^{-2}$) and the thermal effusivity of the substrates obtained from **Section 3.2**.

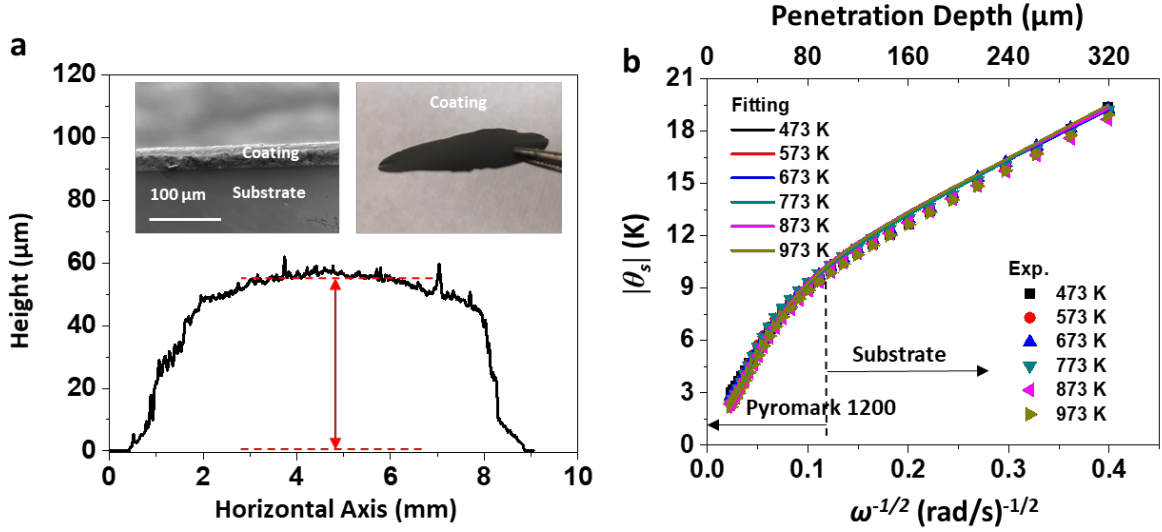

**Figure 7.** MPR measurement of Pyromark 1200 coating on an 8YSZ substrate. (a) Typical surface profile of Pyromark 1200 coating with a surface roughness of ~ 3 μm. The left inset shows the cross-sectional SEM image of the coating and the right inset shows a photograph of the coating after peeled off from the substrate. (b) Thermal response of 53 μm Pyromark 1200 thin-film coating on an 8YSZ substrate from 473 K to 973 K. The penetration depth in the top $x$ -axis is estimated based on the typical thermal diffusivity value of the coating ($\alpha_f = 0.3\ mm^2 s^{-1}$).

**Table 2** shows the obtained thermal effusivity $e_f$ and thermal diffusivity $\alpha_f$ of the Pyromark 1200 coating. The $R^2$ values of fitting the experimental thermal response to **Equation 4** are greater than 0.99 for all the temperatures. The thermal conductivity $k_f$ and the volumetric heat capacity $\rho_f c_f$ of the Pyromark 1200 can be calculated by from the $e_f$ and $\alpha_f$. The extracted thermal conductivity of Pyromark 1200 and black spinel oxide coating is relatively low, 0.4 ~ 0.6 W m⁻¹ K⁻¹ at 300 K, which can be attributed to the high porosity of the coating. The coatings are mainly composed of silica-resin and spinel oxide, whose thermal conductivities are both in the range of 1 ~ 3 W m⁻¹ K⁻¹ (**Figure 6b** and *Supplementary Information Figure S10*). Therefore, the coating is believed to be porous. As shown in *Supplementary Information Section S5*, the measured density of Pyromark 1200 coating is less than 1250 kg m⁻³, which is considerably lower than that of silica resin (~2300 kg m⁻³) and spinel oxide (5000 ~ 7000 kg m⁻³ ). Besides, the volumetric heat capacity of Pyromark 1200 is measured to be ranging from 1166 to 1713 kJ m⁻³ K⁻¹, which is lower than the common values for most fully-dense solids (~ 3000 kJ m$^{-3}$ K$^{-1}$), further confirming that the coating is highly porous. The porosity nature of the coatings is also consistent with earlier microstructure characterizations on similar coatings made by the authors [42, 46]. The volumetric heat capacity for the Pyromark 1200 increases as temperature increases from 473 K to 973 K, which agrees with the trend for glass [36, 47] and spinel oxide (***Supplementary Information Figure S10***).

**Table 2.** Summary of measurement results for 53 µm thick Pyromark 1200 on an 8YSZ substrate.

| $T_o$ (K) | $e_f$ (J s$^{-1/2}$ m$^{-2}$ K$^{-1}$) | $D_f$ (s$^{1/2}$) | $\alpha_f$ (mm$^2$ s$^{-1}$) | $R^2$ | $k_f$ (W m$^{-1}$ K$^{-1}$) | $\rho_f c_f$ (kJ m$^{-3}$ K$^{-1}$) |
|---|---|---|---|---|---|---|
| 473 | 785.1 | 0.0787 | 0.453 | 0.9980 | 0.528 | 1166.5 |
| 573 | 797.6 | 0.0858 | 0.381 | 0.9987 | 0.492 | 1292.2 |
| 673 | 865.5 | 0.0909 | 0.340 | 0.9996 | 0.505 | 1484.3 |
| 773 | 862.2 | 0.0995 | 0.284 | 0.9999 | 0.459 | 1617.9 |
| 873 | 861.5 | 0.0991 | 0.287 | 0.9999 | 0.462 | 1608.1 |
| 973 | 872.0 | 0.1041 | 0.259 | 0.9995 | 0.444 | 1713.4 |

**Figure 8.** shows the MPR measurements of different solar-absorbing coatings, with variation in the coating materials, substrate materials and coating thickness. **Figure 8a** compares the thermal responses of Pyromark 1200 with similar coating thickness but on different substrates at 973 K, i.e., 53 µm thick coating sprayed on 8YSZ and 45 µm thick coating sprayed on SS304. The thermal response curves for other temperatures (from 473 K to 973 K) can be found in the ***Supplementary Information Figure S3***. It is apparent that the transitional region is more distinguishable when using SS304 as the substrate due to the larger contrast in the thermal effusivities between the coating and the substrate, leading to a higher sensitivity to thermal diffusivity of the coating. **Figure 8b** shows the extracted thermal conductivity and thermal diffusivity of Pyromark 1200 from 473 K to 973 K, by fitting the experimental data using **Equation 4**. Thermal conductivities of Pyromark 1200 measured from the two different substrates (SS304 and 8YSZ) are close,

ranging from 0.4 to 0.6 W m$^{-1}$ K$^{-1}$ within the temperature range of 473 K to 973 K. The discrepancy may come from the difference in the thermal expansion coefficients for the two substrates which could lead to microstructural changes, such as cracking in the coating during curing [42-44]. Nevertheless, the deviation of the measured results is still within 15% between the two samples. **Figure 8c** shows the thermal responses of coatings with different thicknesses on SS304 substrates at 973 K. The thermal response curves for other temperatures (from 473 K to 973 K) can be found in the ***Supplementary Information Figure S3***. All the curves show similar slopes in the low frequency range because they all have SS304 as the substrate. However, due to the different coating thickness, the transitional frequency or penetration depth, where the slope starts to change, varies for different samples. The transitional penetration depth increases with the coating thickness, e.g., ~ 80 μm for the 45 μm Pyromark 1200 and ~ 50 μm for the 30 μm Pyromark 1200. It is noted that all the curves converge in the high frequency range which is dominated by the properties of the coatings, indicating similar thermal conductivities of these coatings. As shown in **Figure 8d**, thermal conductivities for the Pyromark 1200 and Pyromark 2500 are determined to range from 0.4 to 0.6 W m$^{-1}$ K$^{-1}$ in the temperature range from 473 K to 973 K. The thermal conductivity for the black spinel oxide coating is slightly higher, ranging from 0.6 to 0.8 W m$^{-1}$ K$^{-1}$ in the same temperature range, and it is also much lower than that the dense pellet made of a similar spinel oxide material (2-2.5 W m$^{-1}$ K$^{-1}$, **Figure 6b**), due to the porous structure of the coating. The low thermal conductivity of the solar-absorbing coatings indicates a possibly considerable temperature drop within the coating under high solar flux. For example, the temperature drop is estimated to be ~ 60 K for a 30 μm thick coating under the heat flux of 1000-sun (***Supplementary Information Figure S9***).

In this work, thermal properties of bulk and coating materials are measured up to 973 K. Higher temperature can be achieved using higher heating power. However, it is important to ensure the temperature uniformity on the sample surface at higher temperature. In addition, the minimum coating thickness is 11 μm in this study because the skin depth of Pyromark coating in IR-spectrum is ~ 5 μm [43, 44]. If the coating was thinner than the skin depth, the IR detector would receive the signal from the substrate, leading to the inaccurate reading. However, if a coating has very shallow IR skin depth (e.g., metallic materials with high IR extinction coefficient), then the MPR technique can be applied to thinner coatings. For instance, Martan et al. [31] were able to measure ~500 nm thick tungsten film because of the very shallow skin depth of tungsten.

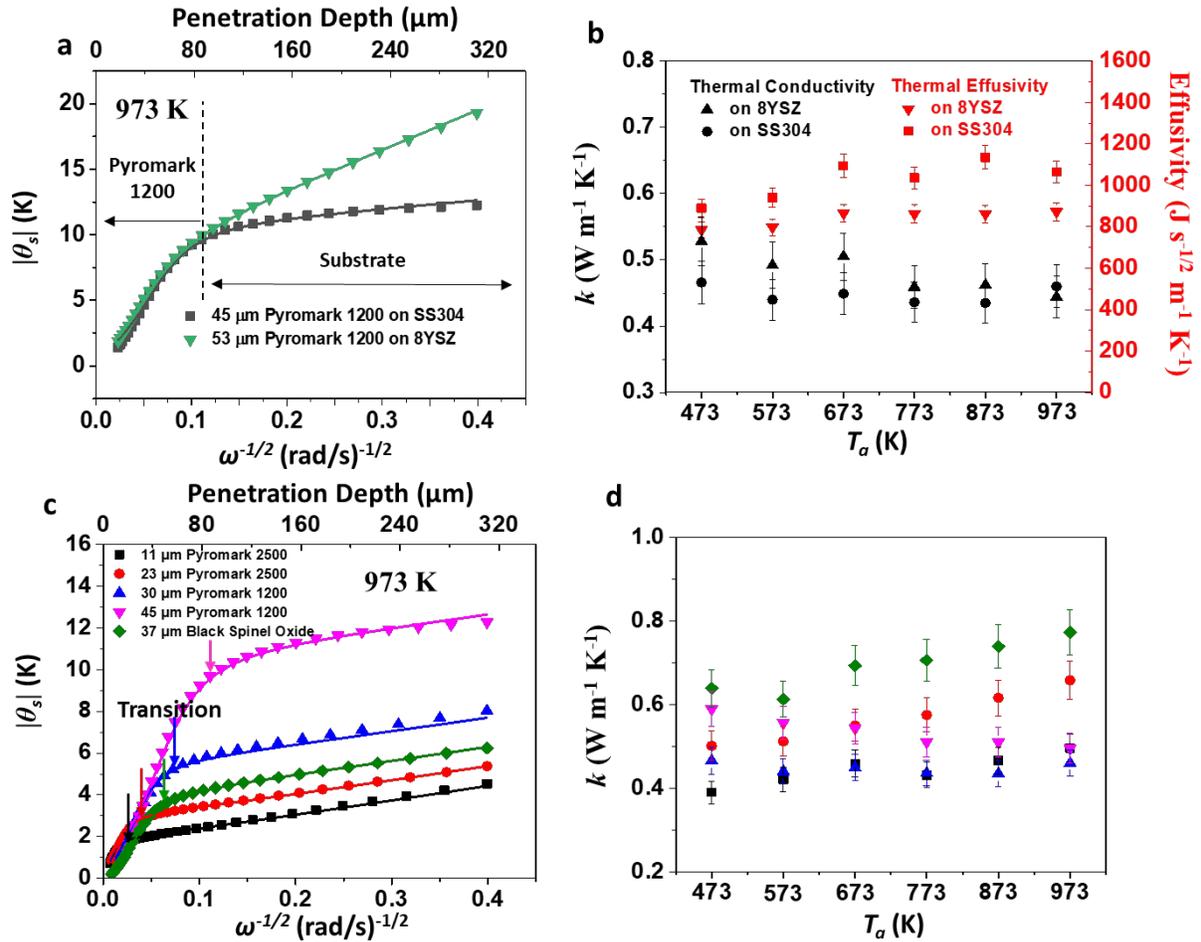

**Figure 8.** Measurement of different solar-absorbing coatings with various coating materials, thicknesses and substrate materials. (a) Effect of substrate materials on the thermal response of the coating measured at $T_o$ = 973 K. (b) Extracted thermal conductivity (left *y*-axis) and thermal effusivity (right *y*-axis) of Pyromark 1200 coated on 8YSZ and SS304 substrates from 473 K to 973 K. (c) Thermal responses of different solar-absorbing coatings with variable thickness spray-coated on SS304 substrates measured at $T_o$ = 973 K. (d)

Thermal conductivity of different solar-absorbing coatings spray coated on SS304 substrates, measured from 473 K to 973 K.

4. Conclusion

In this work, we develop a general high-temperature MPR setup to measure coating materials up to 973 K. The MPR takes the advantage of facile sample preparation on rough surfaces and high sensitivity at high temperatures due to the increasing emissive power. The designed sample holder can ensure the temperature uniformity for high temperature measurement within the infrared detection spot along the lateral direction and the thermal penetration depth of the measurements at high temperatures. The MPR setup is validated with the measurement on bulk materials with thermal conductivities ranging from 1 to 30 W m$^{-1}$ K$^{-1}$ from 473 to 973 K, with the maximum deviation of measured thermal conductivities from the literature or reference values less than 10%. Several types of solar-absorbing coatings are also measured, including Pyromark 1200, Pyromark 2500 and black spinel oxide coatings, with variation in the coating thickness of (10 ~ 50 μm) and substrate materials. Both the thermal effusivity and thermal diffusivity of the coatings are obtained to yield the thermal conductivity and volumetric heat capacity. We find that the thermal conductivity of Pyromark 1200 and Pyromark 2500 ranges from 0.4 to 0.6 W m$^{-1}$ K$^{-1}$ and that for the black spinel oxide coating is ranged from 0.6 to 0.8 W m$^{-1}$ K$^{-1}$, within the temperature window of 473 K to 973 K. The measurement results indicate that it is necessary to consider the thermal resistance of the solar-absorbing coating in the design of CSP solar receivers, because the temperature drop within the coating can be significant. The MPR methodology demonstrated in this work may provide an attractive and convenient thermal conductivity diagnostic tool at high temperature on samples that are otherwise challenging to measure using other techniques, for example, specimens with surface roughness and/or small thickness down to ~10 μm.

**Supplementary Information**

Photograph of MPR system; Thermal response of bulk materials; Thermal response of thin solar-absorbing coatings on substrate; COMSOL simulation of sample holder at high temperature; Characterization of Pyromark 1200 coating; Thermophysical properties of Pyroceram 9606; Calibration of MCT calibration; Estimation of temperature drop within the solar-absorbing coating; Thermophysical properties of spinel oxide; Repeatability test at 973 K; Validation of fitting for coatings.

**Acknowledgements**

This material is based upon work supported by the U.S. Department of Energy's Office of Energy Efficiency and Renewable Energy (EERE) under Solar Energy Technologies Office (SETO) Agreement Number DE-EE0008379. The views expressed herein do not necessarily represent the views of the U.S. Department of Energy or the United States Government.

**Author contribution statement**

RC conceived the idea. RC and PL supervised the project. J.Z. and KC built the MPR setup, performed the modeling measurements, and analyzed the data. YP prepared and characterized the coatings. QW and XW measured the standard bulk samples using the LFA at UCSD and UA respectively. JZ and RC wrote the manuscript, with inputs from all the coauthors.

**Figure and Table Captions**

**Figure 1**. Principle of MPR measurement on a two-layer sample.

**Figure 2.** Details of MPR system. (a) Configuration of MPR system. (b) Photograph (top) and schematic (bottom) of the high-temperature sample holder for the MPR measurement. Samples are placed at the center of the holder made of Inconel 625 alloy.

**Figure 3.** Finite-element simulation using COMSOL for the temperature distribution in the high-temperature sample holder. (a) Temperature profile of the sample holder with the heater power of 170 W and the DC component of laser power of 2 W. The upper row shows the external surface while the lower row shows the cross-section of the holder. (b) Temperature distribution within the sample along the thickness direction. (c) Temperature distribution along the radial direction on the sample surface.

**Figure 4**. Simulation of the thermal response and the sensitivity analysis for a 40 μm Pyromark 1200 coating on a SS304 semi-infinite substrate. (a) $|\theta_s|$ (left y-axis) and slope $m$ of the $|\theta_s|$ vs $\omega^{-1/2}$ curve (right y-axis) as a function of $\omega^{-1/2}$ (bottom x-axis), and the corresponding thermal penetration depth in the coating ($L_p = \sqrt{2\alpha_f/\omega}$, top x-axis)). (b) Phase of thermal response. (c) Sensitivity of $|\theta_s|$ on $e_f$, $\alpha_f$ and $e_s$. (d) Sensitivity of $m$ on $e_f$, $\alpha_f$ and $e_s$.

**Figure 5**. Validation of MPR measurement of bulk materials. (a) Measurement of laser heat flux using Pyroceram 9606 at 473 K as a reference sample. (b) Repeatability of MPR measurement and effect of coating materials. All the measurements were done on Pyroceram 9606 substrates. (c) IR detection signal of MPR measurement on a Pyroceram 9606 sample for sample surface baseline temperature ($T_o$) ranging from 473 K to 973 K. (d) Slope $\tilde{m}$ of $V_{IR}$ vs $\omega^{-1/2}$ as a function of $T_o^3$, following **Equation 10**.

**Figure 6.** Validation of high temperature MPR measurement of bulk materials.(a) Measured thermal effusivity of different bulk materials up to 973 K. (b) Extracted thermal conductivity of different bulk materials up to 973 K by relating to the thermal effusivity using the $\rho c$ values. (c) Comparison of the measured thermal effusivity to that from literatures or known values at $T_0 = 673\ K$. (d) Comparison of the measured thermal conductivity to that from literatures or known values at $T_0 = $

$673\ K$.

**Figure 7.** MPR measurement of Pyromark 1200 coating on a 8YSZ substrate. (a) Typical surface profile of Pyromark 1200 coating with a surface roughness of ~ 3 μm. The left inset shows the cross-sectional SEM image of the coating and the right inset shows a photograph of the coating after peeled off from the substrate. (b) Thermal response of 53 μm Pyromark 1200 thin-film coating on a 8YSZ substrate from 473 K to 973 K. The penetration depth in the top $x$-axis is estimated based on the typical thermal diffusivity value of the coating ($\alpha_f = 0.3\ mm^2 s^{-1}$).

**Figure 8.** Measurement of different solar-absorbing coatings with various coating materials, thicknesses and substrate materials. (a) Effect of substrate materials on the thermal response of the coating measured at $T_o = 973$ K. (b) Extracted thermal conductivity (left $y$-axis) and thermal effusivity (right $y$-axis) of Pyromark 1200 coated on 8YSZ and SS304 substrates from 473 K to 973 K. (c) Thermal responses of different solar-absorbing coatings with variable thickness spray-coated on SS304 substrates measured at $T_o = 973$ K. (d) Thermal conductivity of different solar-absorbing coatings spray coated on SS304 substrates, measured from 473 K to 973 K.

**Table.1.** Random Error of MPR measurement on Pyroceram 9606 with different coatings

**Table 2.** Summary of measurement results for 53 μm thick Pyromark 1200 on an 8YSZ substrate.